\documentclass[aps,prl,twocolumn,groupedaddress]{revtex4-1}
\usepackage{graphicx}
\usepackage{bm}
\usepackage{mathtools}
\usepackage{amsmath,amssymb}

\newcommand{\CBST}{Cr$_{\rm{x}}$(Bi,Sb)$_{\rm{{2-x}}}$Te$_3$}

\newcommand{\Vg}{$V_{\rm{G}}$}

\newcommand{\rhoxx}{$\rho_{\rm{{xx}}}$}
\newcommand{\rhoxy}{$\rho_{\rm{{xy}}}$}
\newcommand{\sigmaxy}{$\sigma_{\rm{{xy}}}$}
\renewcommand{\vec}[1]{\mbox{\boldmath$1$}}

\newcounter{lastnote}

\def\bc{\begin{center}}
\def\ec{\end{center}}
\def\be{\begin{equation}}
\def\ee{\end{equation}}
\renewcommand{\vec}[1]{\mbox{\boldmath$1$}}

\begin{document}
\title{Giant Anisotropic Magnetoresistance in a Quantum Anomalous Hall Insulator}
\author{A. Kandala}
\author{A. Richardella}
\author{S. Kempinger}
\author {C-X. Liu} 
\author{N. Samarth}
\email{nsamarth@psu.edu}
\affiliation{Department of Physics, The Pennsylvania State University, University Park, Pennsylvania 16802-6300, USA}
 \begin{abstract}
 When a three-dimensional (3D) ferromagnetic topological insulator thin film is magnetized out-of-plane, conduction ideally occurs through dissipation-less, one-dimensional (1D) chiral  states that are characterized by a quantized, zero-field Hall conductance. The recent realization of this phenomenon -- the quantum anomalous Hall effect -- provides a conceptually new platform for studies of edge-state transport, distinct from the more extensively studied integer and fractional quantum Hall effects that arise from Landau level formation. An important question arises in this context: how do these 1D edge states evolve as the magnetization is changed from out-of-plane to in-plane? We examine this question by studying the field-tilt driven crossover from predominantly edge state transport to diffusive transport in \CBST~ thin films, as the system transitions from a quantum anomalous Hall insulator to a gapless, ferromagnetic topological insulator. The crossover manifests itself in a giant, electrically tunable anisotropic magnetoresistance that we explain using the Landauer-B\"uttiker formalism. Our methodology provides a useful means of quantifying edge state contributions to transport in temperature and chemical potential regimes far from perfect quantization.
 \end{abstract}
 
\date{\today}

\maketitle

In the absence of magnetic doping, surface electrons in 3D topological insulators such as (Bi,Sb)$_2$Te$_3$ are helical Dirac fermions protected from back scattering by time reversal symmetry \cite{Hasan2010,Qi2011}. When such a 3D topological insulator is made ferromagnetic by doping with magnetic ions \cite{Liu2012,Checkelsky2012,Zhang2012a,Chang2013,Lee2014,Checkelsky2014,Kou2014} or by integration into magnetic heterostructures\cite{Wei2013,Kandala2013b,Lang2014}, time-reversal symmetry is broken. If the magnetization $M$ is perpendicular to the surface, the dispersion is gapped at the Dirac point\cite{Qi2008,Xu2012}, while $M$ parallel to the surface is expected to keep the Dirac point intact with the origin shifting in k-space. In a ferromagnetic topological insulator thin film with out-of-plane magnetic anisotropy\cite{Yu2010}, the two surfaces are gapped, albeit with opposite mass. The mass domain wall created along the side walls of films then hosts a massless, one-dimensional, dissipation free edge state. If the chemical potential is tuned inside the magnetic gap of the surface states, transport is solely restricted to these 1D edge states, and characterized by a quantized Hall resistivity $\rho_{\rm{xy}} =h/{e^2}$ and (ideally) vanishing longitudinal resistivity $\rho_{\rm{xx}} = 0$. This is known as the quantum anomalous Hall (QAH) state and was recently realized experimentally in thin films of \CBST{} \cite{Chang2013,Checkelsky2014,Kou2014} at dilution fridge temperatures. 

Anisotropic magneto-resistance (AMR)\cite{Thomson1856,Mcguire1975} is the response of the longitudinal resistance of a ferromagnet to the angle between the magnetization and direction of current flow. In this work, by tilting the magnetization of QAH insulator in-plane, we observe a novel, giant AMR effect that serves as a powerful quantitative probe of edge transport, even in temperature and chemical potential regimes away from perfect quantization (for instance, at high temperatures compared to the energy scale of the magnetic gap). Tilting the magnetization in-plane closes the magnetic gap and consequently destroys the edge states, leading to a crossover from dissipation-less to diffusive transport. We demonstrate that this crossover yields a sharp change in the longitudinal resistivity (\rhoxx{}) whose angular dependence is unlike classical AMR and show that modeling this dependence can be used to quantify edge state contributions to transport, and the extent of its inter-mixing with dissipative channels.

The magnetic topological insulator devices used in this study were fabricated from 8 quintuple layer thick epitaxial films of \CBST{} grown by molecular beam epitaxy on $(111)$ SrTiO$_3$ (STO) substrates and measured in a vector magnet cryostat down to 280 mK (Figure 1a, see Methods for details). The large dielectric constant of the STO substrate at cryogenic temperatures allows effective electrical back gating for tuning the position of the chemical potential\cite{Chen2010}.Three samples were studied with varying proximity to the QAH state (Table 1). In sample A, even at 280 mK, the AHE takes a maximum value \rhoxy $\sim 0.95h/e^2$  (\sigmaxy $\sim 0.99h/e^2$) and the longitudinal resitivity its minimum value \rhoxx $\sim 0.25h/e^2$, both  at zero field, at charge neutrality (\Vg$^{0}=-100$V). This non-vanishing \rhoxx, accompanied by \rhoxy~ taking near-quantized values, is indicative of dissipative channels co-existing with the 1D chiral edge modes. \rhoxx{} also displays a prominent hysteretic peak at magnetization reversal, and takes very large values approaching 260\%. This large effect has been interpreted as enhanced backscattering between edge channels via domain walls at the coercive field, in a manner similar to the sharp \rhoxx~ rise in the quantum Hall effect as scattering channels are populated (Figure 1b,c). Below 3K near charge neutrality (\Vg=\Vg$^{0}$), \rhoxx{} displays a metallic behavior as \rhoxy{} approaches quantization (Figure 1d,e). This metallic behavior is another signature of edge state transport and \rhoxx{} recovers its insulating behavior when the chemical potential is tuned far out of the magnetic gap (see Supplementary information). Similarly, in the gate voltage dependence, as we tune the chemical potential towards the magnetic gap, \rhoxx{} drops sharply, as \rhoxy{} rises and finally flattens close to its maximum value. 

\begin{figure}
\includegraphics[width=3.5in]{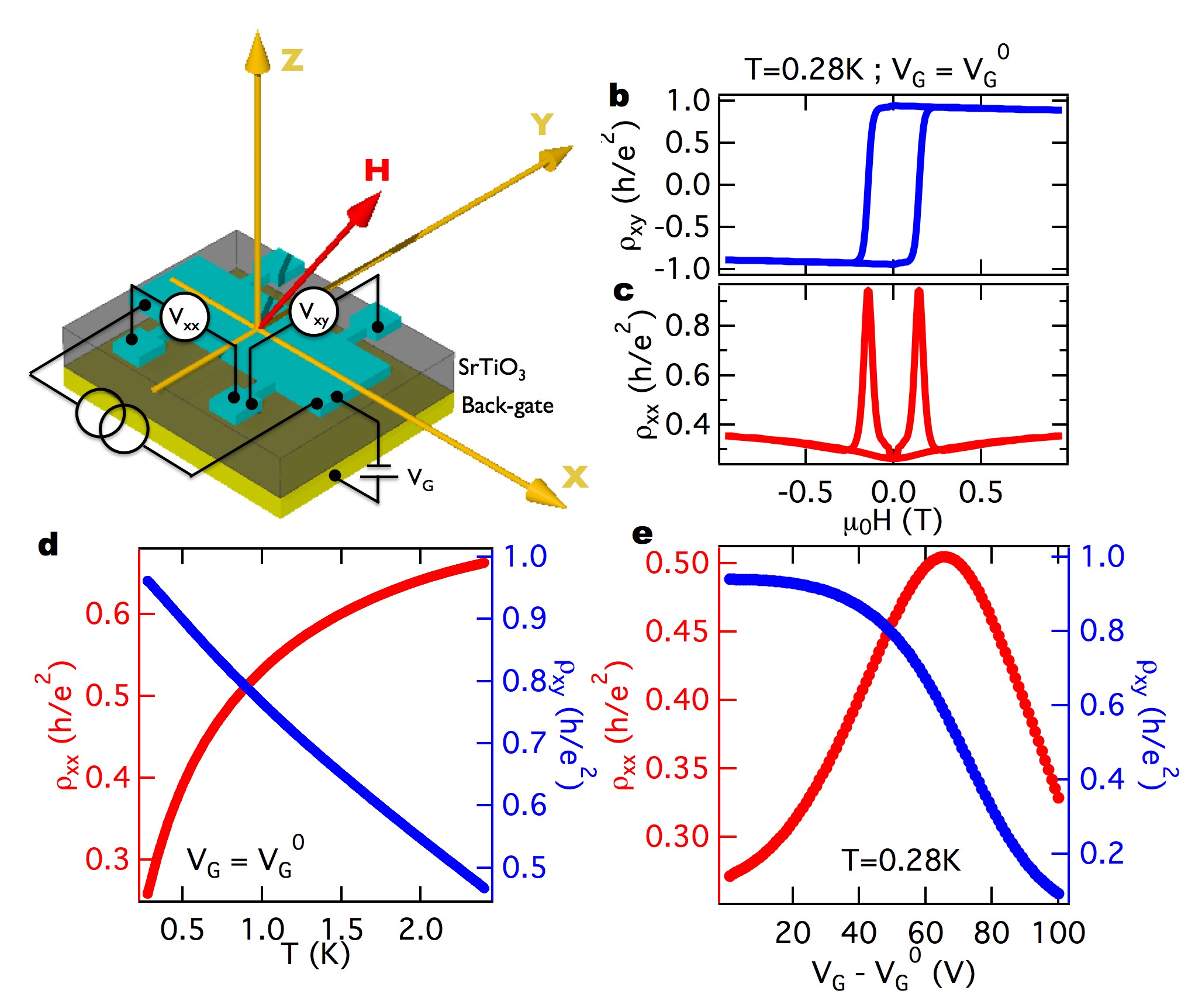} 
\caption{ (a) Cartoon of device and measurement scheme. Current flows in the $x$ direction, and $z$ is the direction perpendicular to plane. (b) Near-quantized anomalous Hall effect at charge neutrality (\Vg$^{0}= -100$V) and $T = 280$ mK, with \rhoxy~ taking a maximum value of 0.95 $h/e^2$ at zero-field. (c) Corresponding longitudinal magneto-resistance, displaying a minimum \rhoxx~ of 0.25 $h/e^2$, also at zero-field. (d) and (e) Characteristic signatures of quantum anomalous Hall effect in the temperature and gate voltage dependence, respectively, of \rhoxx{} (red) and \rhoxy~ (blue) at zero-field (after magnetic training). As edge-state transport dominates over diffusive transport, the decrease in \rhoxx{} is complemented by a rise in \rhoxy{}.}
\label{Fig1}
\end{figure}

\begin{table}
\centering
\caption{Comparison of transport properties and AMR fitting parameters of 3 different \CBST~ thin film devices with varying degrees of dissipation.} 
  \centering
    \begin{tabular}{|c c c c c c|}
    \hline
Device & \rhoxy $(h/e^2)$ & \rhoxx$(0) (h/e^2)$ & $\eta$ & $R_{d}^{-1}$ &AMR (\%)\\ \hline
   A & 0.95 & 0.25 & 0.66 &  0.62 & 140.9\\
   B & 0.72 & 0.55 & 0.16 & 0.54 &  75.6\\ 
   C & 0.48 & 2.79 & 0.013 & 0.025 & 100.2 \\ \hline
\end{tabular}
\end{table}

Having established the signatures of edge state transport, we now examine their evolution as the magnetization is tilted in-plane by an external 1T radial field (Figure 2). The temperature dependence of \rhoxx{} in Figure 2a shows a metal-insulator transition as the magnetization is tilted from out-of-plane to in-plane, irrespective of the azimuthal angle. Similarly, in the gate voltage dependence (Figure 2b), as the chemical potential is tuned into the magnetic gap, the drop in \rhoxx{} is replaced by a sharp rise when the magnetization is in-plane. This is indicative of a crossover from dominant 1D edge transport to diffusive transport, as the system transitions from a QAH regime to a gapless ferromagnetic topological insulator (when magnetized in-plane). Figures 2a and 2b already reveal a ``giant" AMR effect when the chemical potential is in the magnetic gap, and this crossover is the origin of the novel angular dependence of the AMR that we now discuss.

\begin{figure*}
\includegraphics[width=6.8in]{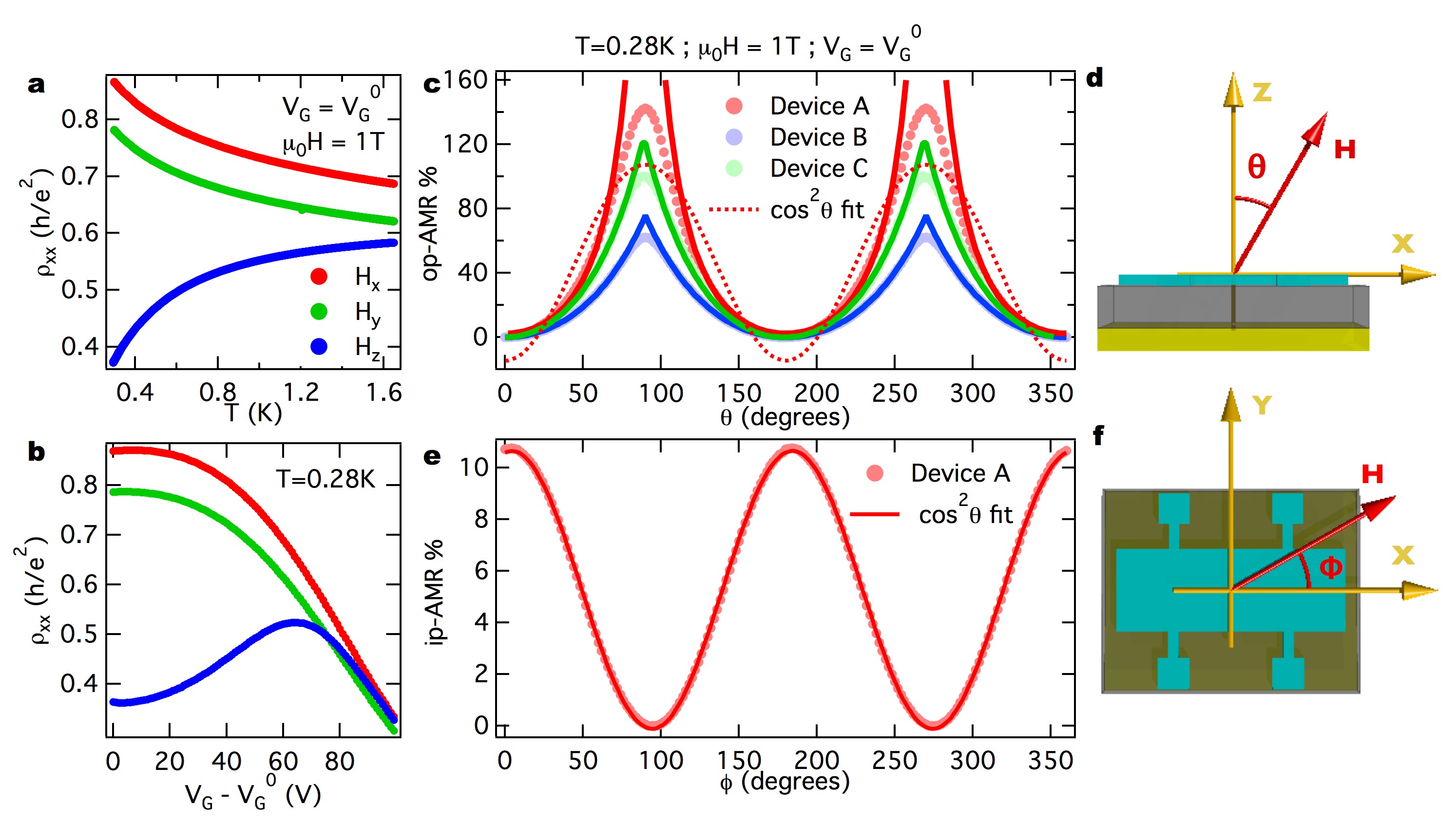} 
\caption{An external radial field of 1 T is used for all the data in this figure. (a) Transition from metallic to insulating behavior in the temperature dependence of $\rho_{xx}$ at \Vg~ = \Vg$^{0}$, as the magnetization is tilted from out-of-plane to in-plane. The red, green and blue curves correspond to the $x$, $y$, and $z$ directions respectively. (b) Gate voltage dependence of $\rho_{xx}$ at $T =$ 280 mK when the film is magnetized along the $x$ (red),$y$ (green) and $z$ (blue) directions. (c) Out-of-plane anisotropic magneto-resistance of device A (light red, circles) at charge neutrality, as the field is rotated in the $x-z$ plane. The angular dependence is unexplained by a conventional $\cos^2\theta$ fit (red, dotted line) and is understood by fits (red, bold line) that use an edge-dissipation mixing model. AMR data (circles) and corresponding fits (bold lines) from device B (light blue) and device C (light green), described in the text. (d) Cartoon of the device geometry highlighting field rotation from out-of-plane to in-plane, in the $x-z$ plane. (e) In-plane anisotropic magneto-resistance of device A (light red, circles) at charge neutrality, as the field is rotated in the $x-y$ plane. The angular dependence is explained by a typical $\cos^2\phi$ fit (red, bold line). (f) Cartoon of the device geometry highlighting field rotation in the $x-y$ plane.}
\label{Fig2}
\end{figure*}

We systematically probe the crossover from edge-dominated to diffusive transport as we rotate a 1 T external magnetic field from perpendicular-to-plane to in-plane. For a 1 T field, the AMR, defined as $(\rho_{\rm{xx,max}}-\rho_{\rm{xx,min}})/\rho_{\rm{xx,min}}$ takes a maximum value $\sim$ 140\%, and its angular dependence is completely inconsistent with the well-known $\cos^2\theta$ angular dependence seen in conventional ferromagnets. To explain the observations, we model the angular dependence using a simplified four-terminal Landauer-B\"uttiker formalism $I_{i}=\Sigma_{j}G_{ij}(V_{i}-V_{j})$ that assumes the coexistence of both dissipation-less edge channels and dissipative bulk and surface channels. The dissipative channels likely arise from the gapless, two-dimensional (2D) surface Dirac states of the side walls when the chemical potential is in the magnetic gap\cite{Wang2013}. We note that these surface states are not gapped because the external magnetic field and the magnetization are always parallel to the plane of the side walls. We cannot exclude other possible origins of dissipation, such as the bulk bands, mid-gap impurity bands, and even the gapped surface states. These details are however not important for the model. The conductance $G_{ij}$ is given by $G_{ij}=T_{ij}(h/e^{2})$ where the transmission co-efficient $T_{ij}$ takes the form $T_{ij} = \eta\delta_{i,j+1} + t_{ij}$. Here, $\eta$ is the contribution from edge modes that only transmit between leads $j$ and $j+1$, and $t_{ij}$ corresponds to dissipative contributions. From calculations explained in detail in the supplementary section, we derive the following expression for the magneto-resistance:
\begin{equation}
R=\frac{h}{e^2}\frac{1}{\eta (2 + R_{d}\eta) + R_{d}^{-1}},
\end{equation}
where $R_{d}$ is a measure of the ``dissipative'' resistance (and $R_{d}^{-1}$ is the dissipative transmission coefficient). Physically, the first term in the denominator is the conductance of an inter-mixed channel, that constitues the edge states scattering into dissipative pathways. The second term though corresponds solely to the dissipative channels. The transmission coefficient for the edge modes $\eta$ depends on the gap $\Delta$ induced by the magnetization. We expect $\eta$ to be suppressed from its maximum value of 1 as the magnetic gap is reduced $\Delta$: this arises from enhanced inter-mixing with dissipative channels that accompanies an increase in penetration length $\lambda$ of the edge-state wave functions, where $\lambda =\hbar v_{F}/\Delta$ where $v_{F}$ is the Fermi velocity. We thus assume a phenomenological expression for the edge state transmission as:
\begin{equation}
\eta=1-e^{-l_{0}/\lambda}=1-e^{-\Delta/\Delta_{0}}
\end{equation}
where $l_{0}$ is a length scale that quantifies the strength of the inter-mixing. This expression satisfies the physical requirement at the two extrema: $\Delta \to 0 \implies \eta \to 0$ and $\Delta \to \infty \implies \eta \to 1$. The angle dependence of the magneto-resistance then arises via the dependence of the magnetic gap on the perpendicular component of field, i.e. $\Delta=\Delta_{m}\cos\theta$. $\Delta_{m}$ is the size of the magnetic gap, when the film is magnetized out-of-plane. The physics behind the model is depicted in Figure 3. We now use Eq. 1 to fit our angular data with two free parameters: $R_{d}$ and $\Delta '$ (defined as $\Delta ' =\Delta_{0}/\Delta_{m}$). The model clearly fits the data very well, up to tilt angles $\sim 70^{o}$, as seen in Figure 2b, confirming that the novel, giant AMR effect has its origins in the edge states. The model deviates at large tilt angles when the field approaches the in-plane orientation. In this regime, the dominant resistance is associated with the dissipative channels, whose angular magneto-resistance is not accounted for in our model ($R_{d}$ is treated as a constant).

\begin{figure}
\includegraphics[width=3in]{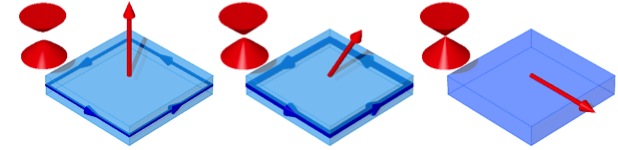} 
\caption{A pictorial representation of the physical origins of the giant AMR effect. The penetration length of edge states (depicted in dark blue) is dependent on the size of the magnetic gap $\Delta$, which in turn depends on the out-of-plane component of magnetization. As the external field is tilted in-plane, the size of the magnetic gap decreases (illustrated by the energy-momentum dispersions, in red). The penetration length of the edge states into the bulk (depicted in light blue) increases, as indicated in the middle panel, eventually leading to complete mixing as the magnetization is tilted in-plane (last panel).} 
\label{Fig2}
\end{figure}

Beyond explaining the angular dependence of this novel AMR, our model also presents a useful approach to quantify edge state contributions to transport in an edge-dissipation ``mixed" QAH insulator. The fitting parameter $\Delta '$ can be used to extract the edge transmission coefficient $\eta = 1 - e^{-1/\Delta '}$. This is used to quantify and compare edge transport contributions in two other samples with varying degrees of dissipation, whose AMR is also shown in Figure 2c. The strength of the dissipation is inferred by comparing the zero-field \rhoxx~ and the magnitude of the AHE in these films. The angular dependence and fits to data from devices B and C also serves as a test of reproducibility for the giant AMR. Table 1 summarizes the different device characteristics and the extracted fitting parameters, all at charge neutrality and at the same temperature. The extracted values of $\eta$ consistently characterize the degree to which the samples are close to the ideal QAH regime. Also, it is important to point out that the magnitude of the AMR effect consistently depends on the relative values of $\eta$ and $R_{d}^{-1}$ (Eq. 1).

As we mentioned earlier, conventional ferromagnets show an AMR described by a $\cos^2\phi$ angular dependence. This behavior is indeed observed for in-plane measurements (Fig. 2e). Compared to the giant AMR, the in-plane AMR only has magnitude of  $\sim$ 10.7\%, at charge neutrality (Figure 2e), and also shows a far weaker temperature dependence (difference between upper two curves in Fig. 2a). With the field and magnetization in-plane, the sample is expected to be in a diffusive regime, since tuning the surface states gapless destroys the edge states. More detailed mapping of the in-plane behavior is presented in the supplementary section.

Finally, we present a more detailed picture of dependence of the giant AMR on the gate voltage and temperature. This is shown for device A in Figure 4. The effect is electrically tunable over an order of magnitude, and is largest in magnitude with the chemical potential in the magnetic gap, when edge state contributions to transport are maximum. As the chemical potential is tuned out of the magnetic gap, surface state and bulk dissipative channels are populated, reducing edge state contributions and consequently the magnitude of the AMR. This is captured in the gate dependence of the edge state transmission coefficient $\eta$ (Fig. 4b), extracted from our fits to the data of Fig. 4a. The qualitative behavior of $R_{d}^{-1}$ is also reasonable, since the dissipative transmission coefficient increases as more electrons are accumulated. Similarly, the temperature dependence of the AMR (Fig. 4c) also captures the competition between the edge states and dissipative channels since  our model quantifies edge transport in temperature regimes away from perfect quantization (Fig. 4d). While perfect Hall quantization has required dilution fridge temperatures, our results here demonstrate that the temperature scale for observing this novel AMR, with its origins in the edge-states, can be significantly higher. Furthermore, both the gate and temperature dependence of the fitting parameters reveal the interplay between the edge states and dissipation channels: an increase in dissipative transmission is accompanied by a reduction in edge transmission, and vice versa.

\begin{figure}
\includegraphics[width=3.5in]{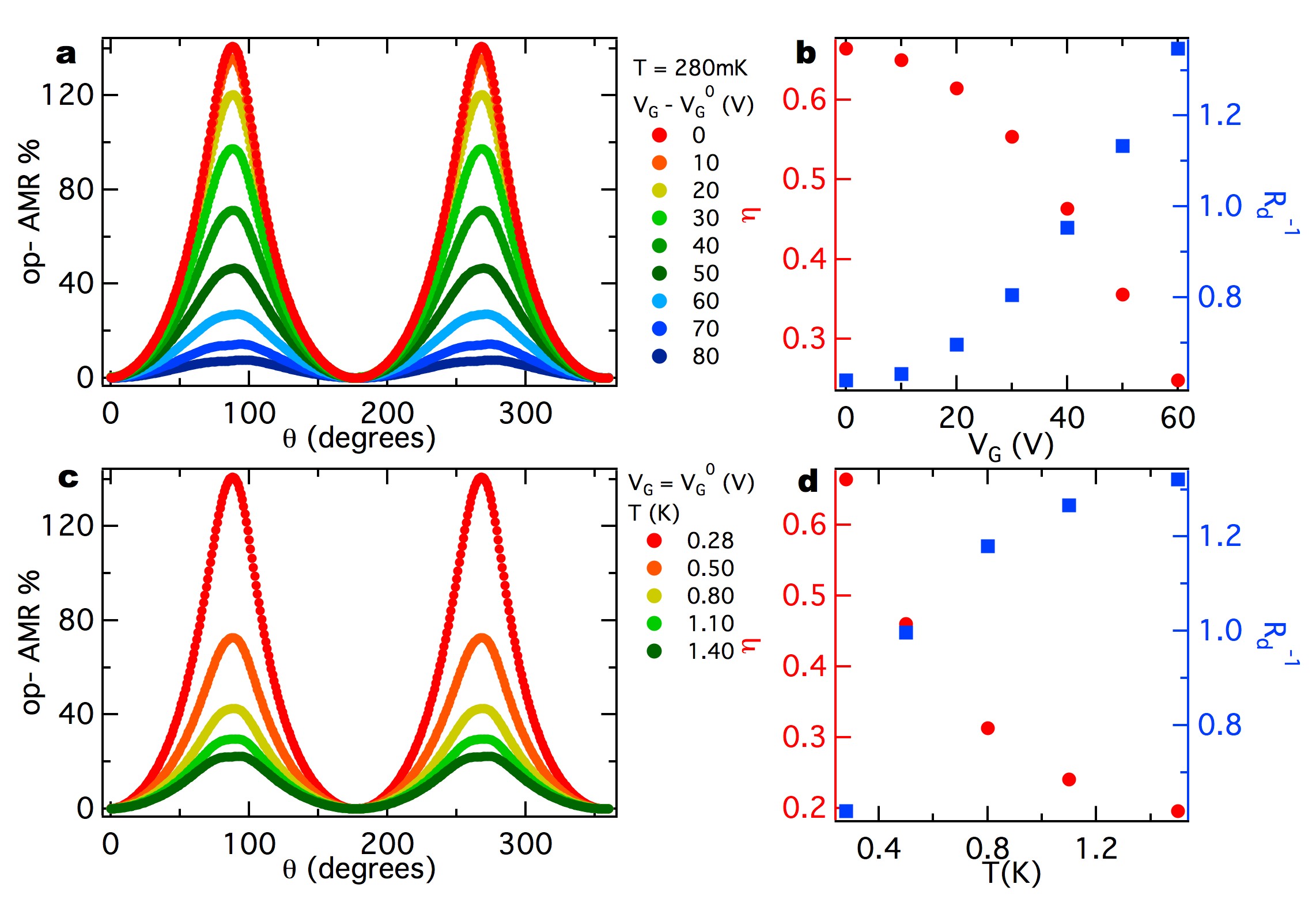} 
\caption { (a) Out-of-plane AMR at $T = 280$ mK for a range of gate voltages. The AMR takes maximum values when the film is tuned to charge neutrality. (b) Gate voltage dependence of the edge-state transmission coefficient $\eta$ (red) and the dissipative transmission coefficient $R_d^{-1}$ (blue) at 280 mK, extracted from fits to the data of (a).  (c) Out-of-plane AMR at \Vg$^{0}$ for a range of temperatures. (d) Temperature dependence of $\eta$ (red) and $R_d^{-1}$ (blue) at \Vg~ = \Vg$^{0}$ extracted from fits to the data of (c).} 
\label{Fig3}
\end{figure}

In summary, we have probed the evolution of edge states in a QAH insulator, as its magnetization is tilted in-plane, both by angular magneto-resistance measurements and by developing a model based on the Landauer-B\"uttiker formalism that accounts for co-existing dissipative channels. We demonstrate a crossover from dominant 1D edge transport to diffusive transport driven by the closing of the surface state magnetic gap. This crossover results in a novel, gate-tunable, giant AMR effect that serves as a quantitative probe of the interplay between the chiral edge modes and dissipative channels, even in non-ideal QAH regimes. This methodology is relevant to ongoing efforts to raise the temperature scale of the quantum anomalous Hall effect \cite{Chang2015}. Furthermore,  as with recent advances that exploit the interplay between spin polarized states in topological insulators and ferromagnetism \cite{Mellnik2014,Fan2014,Li2014}, the observation of a giant AMR effect may provide a useful basis for proof-of-concept devices for ``topological spintronics."

{\bf Methods}

{\it Material synthesis and characterization.}   The \CBST~ thin films studied in this work were grown under ultra-high vacuum (low $10^{-10}$ Torr) by molecular beam epitaxy on $(111)$ STO . The source material was high purity Cr, Bi, Sb and Te thermally evaporated from conventional Knudsen cells. For device C, e-beam evaporation of Cr was employed instead. Prior to growth, the STO substrates were subjected to an ex-situ high temperature (925 \textsuperscript{o}C - 950 \textsuperscript{o}C) anneal in an oxygen atmosphere to obtain a smoother substrate morphology. The topological insulator film of device A was capped {\it in-situ} with a thin layer of room temperature deposited Al, which naturally oxidized upon exposure to atmosphere. Devices B and C were uncapped. The thickness of the topological insulator films was estimated by comparison with the growth rate of calibration samples, whose thickness was measured using high resolution transmission electron microscopy and also confirmed by atomic force microscopy of the Hall bar step edge. 

{\it Transport measurements} All the devices used in this study were fabricated by mechanically scratching MBE-grown films into Hall bar geometries with typical lateral dimensions of around 1 mm. For example, device A has channel dimensions 1 mm $\times$ 0.5 mm.  Since these Hall bars were mechanically scratched, the raw data was symmetrized to account for geometric offsets. The angular measurements were performed in an Oxford Triton He$^3$ vector magnet system with a radial field limit of 1 T and a base temperature of 280 mK. All transport measurements typically employed standard low frequency (19 Hz) lock-in techniques with source currents in the range 10-20 nA, and a Keithley 6430 source meter for back gating. 

{\bf Acknowledgements}

The authors acknowledge support from DARPA MESO (Grant No. N66001-11-1-4110), ONR (Grant No. N00014-12-1-0116-0117) and ARO MURI (Grant No. W911NF-12-1-0461). We acknowledge use of the NSF National Nanofabrication Users Network Facility at Penn State.

\bibliography{Bibliography}
\end{document}